\documentclass[twoside,twocolumn,9pt]{article}
\usepackage{extsizes}
\usepackage[super,sort&compress,comma]{natbib} 
\usepackage[version=3]{mhchem}
\usepackage[left=1.5cm, right=1.5cm, top=1.785cm, bottom=2.0cm]{geometry}
\usepackage{balance}
\usepackage{mathptmx}
\usepackage{sectsty}
\usepackage{graphicx} 
\usepackage{lastpage}
\usepackage[format=plain,justification=justified,singlelinecheck=false,font={stretch=1.125,small,sf},labelfont=bf,labelsep=space]{caption}
\usepackage{float,dblfloatfix}
\usepackage{fancyhdr}
\usepackage{fnpos}
\usepackage{nccmath, mathtools}
\usepackage[english]{babel}
\addto{\captionsenglish}{
}
\usepackage{array}
\usepackage{droidsans}
\usepackage{charter}
\usepackage[T1]{fontenc}
\usepackage[usenames,dvipsnames,table,xcdraw]{xcolor}
\usepackage{setspace}
\usepackage[compact]{titlesec}
\usepackage[hidelinks]{hyperref}
\usepackage[make-links=true]{acro}
\usepackage{enumerate,enumitem}
\usepackage{multirow, booktabs}
\definecolor{cream}{RGB}{222,217,201}
\usepackage{breakurl}

\usepackage{titling}
\title{Cryptorefills Labs}

\begin{document}

\pagestyle{fancy}
\thispagestyle{plain}
\fancypagestyle{plain}{
\renewcommand{\headrulewidth}{0pt}
}

\makeFNbottom
\makeatletter
\renewcommand\LARGE{\@setfontsize\LARGE{15pt}{17}}
\renewcommand\Large{\@setfontsize\Large{12pt}{14}}
\renewcommand\large{\@setfontsize\large{10pt}{12}}
\renewcommand\footnotesize{\@setfontsize\footnotesize{7pt}{10}}
\makeatother

\renewcommand{\thefootnote}{\fnsymbol{footnote}}
\renewcommand\footnoterule{\vspace*{1pt}%
\color{cream}\hrule width 3.5in height 0.4pt \color{black}\vspace*{5pt}} 
\setcounter{secnumdepth}{5}

\makeatletter 
\renewcommand\@biblabel[1]{#1}            
\renewcommand\@makefntext[1]%
{\noindent\makebox[0pt][r]{\@thefnmark\,}#1}
\makeatother 
\renewcommand{\figurename}{\small{Fig.}~}
\sectionfont{\sffamily\Large}
\subsectionfont{\normalsize}
\subsubsectionfont{\bf}
\setstretch{1.125} 
\setlength{\skip\footins}{0.8cm}
\setlength{\footnotesep}{0.25cm}
\setlength{\jot}{10pt}
\titlespacing*{\section}{0pt}{4pt}{4pt}
\titlespacing*{\subsection}{0pt}{15pt}{1pt}

\fancyfoot{}
\fancyfoot[LO,RE]{\vspace{-7.1pt}\includegraphics[height=9pt]{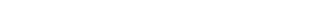}}
\fancyfoot[CO]{\vspace{-7.1pt}\hspace{12.8cm}{\small Cryptorefills Labs, v1.0, 10/2023}}
\fancyfoot[CE]{\vspace{-7.8pt}\hspace{-12.7cm}{\small Cryptorefills Labs, v1.0, 10/2023}}
\fancyfoot[RO]{\footnotesize{\sffamily{~\textbar  \hspace{2pt}\thepage}}}
\fancyfoot[LE]{\footnotesize{\sffamily{\thepage~\textbar\hspace{3.45cm}}}}
\fancyhead{}
\renewcommand{\headrulewidth}{0pt} 
\renewcommand{\footrulewidth}{0pt}
\setlength{\arrayrulewidth}{1pt}
\setlength{\columnsep}{6.5mm}
\setlength\bibsep{1pt}

\makeatletter 
\newlength{\figrulesep} 
\setlength{\figrulesep}{0.5\textfloatsep} 
\newcommand{\topfigrule}{\vspace*{-1pt}%
\noindent{\color{cream}\rule[-\figrulesep]{\columnwidth}{1.5pt}} }
\newcommand{\botfigrule}{\vspace*{-2pt}%
\noindent{\color{cream}\rule[\figrulesep]{\columnwidth}{1.5pt}} }
\newcommand{\dblfigrule}{\vspace*{-1pt}%
\noindent{\color{cream}\rule[-\figrulesep]{\textwidth}{1.5pt}} }
\makeatother

\twocolumn[
  \begin{@twocolumnfalse}
    {\Huge \thetitle \hfill \raisebox{0pt}[0pt][0pt] {\includegraphics[height=55pt]{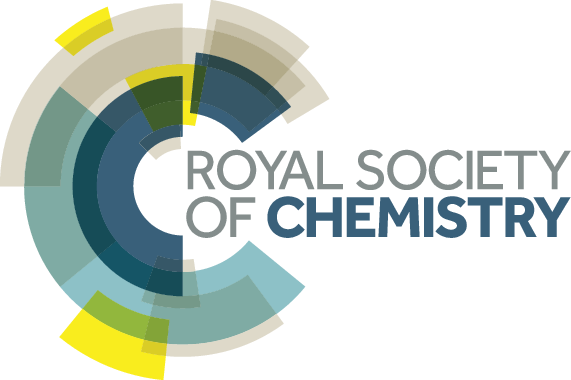}}\\[1ex]
    \includegraphics[width=18.5cm]{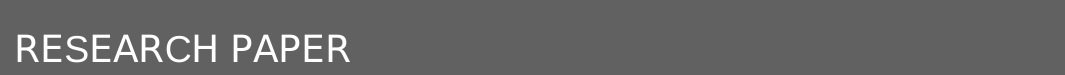}}\par
    \vspace{2.55em}
    \sffamily
    \hspace{-0.1cm}\begin{tabular}{m{4cm} p{14cm} }
    
      \hspace{-0.3cm}\begin{tabular}{l}\vspace{1.25em}\\
            \footnotesize Version: 1.0  \\
            \footnotesize Date: 04 Oct 2023 \\
            \footnotesize DOI: 10.48550/arXiv.2310.02911 \\
        \end{tabular}
        
        & \noindent\LARGE{\textbf{Deciphering the Crypto-shopper: Knowledge and Preferences of Consumers Using Cryptocurrencies for Purchases}} \\ 
        \vspace{0.3cm} & \vspace{0.3cm} \\
        
        \hspace{-0.3cm}\begin{tabular}{l}\footnotesize Publisher: Cryptorefills Labs \\ 
        \footnotesize  License: CC-BY-NC-ND 4.0 \\ \end{tabular} & \noindent\large{Massimiliano Silenzi\textsuperscript{1}, Umut Can Çabuk\textsuperscript{2}, Enis Karaarslan\textsuperscript{3}, Ömer Aydın\textsuperscript{4}.} \\\\

        \vspace{0.25cm}\hspace{-0.34cm}\begin{tabular}{l}\\ \footnotesize Contact: ucabuk@sdsu.edu\\
        \footnotesize\textsuperscript{1}Cryptorefills, The Netherlands\\ \footnotesize\textsuperscript{2}San Diego State University, USA\\ \footnotesize\textsuperscript{3}Muğla Sıtkı Koçman Univ., Turkey\\ \footnotesize\textsuperscript{4}Manisa Celal Bayar Univ., Turkey\end{tabular} & \noindent\normalsize{\textbf{Abstract.} The fast-growing cryptocurrency sector presents both challenges and opportunities for businesses and consumers alike. This study investigates the knowledge, expertise, and buying habits of people who shop using cryptocurrencies. Our survey of 516 participants shows that knowledge levels vary from beginners to experts. Interestingly, a segment of respondents, nearly 30\%, showed high purchase frequency despite their limited knowledge. Regression analyses indicated that while domain knowledge plays a role, it only accounts for 11.6\% of the factors affecting purchasing frequency. A K-means cluster analysis further segmented the respondents into three distinct groups, each having unique knowledge levels and purchasing tendencies. These results challenge the conventional idea linking extensive knowledge to increased cryptocurrency usage, suggesting other factors at play. Understanding this varying crypto-shopper demographic is pivotal for businesses, emphasizing the need for tailored strategies and user-friendly experiences. This study offers insights into current crypto-shopping behaviors and discusses future research exploring the broader impacts and potential shifts in the crypto-consumer landscape. \newline \newline \textbf{Keywords.} Behavioral Analysis, Bitcoin, Blockchain, Cryptocurrencies, Crypto-Shopping, Consumer Expertise, Consumer Segmentation, DeFi, Technology Adoption.} \\
    
    \end{tabular}
 \end{@twocolumnfalse}
\vspace{1cm}
]

\renewcommand*\rmdefault{bch}\normalfont\upshape
\rmfamily
\section*{}
\vspace{-1cm}




\setcounter{tocdepth}{2}  
{\small \balance \tableofcontents } \newpage

\section{Introduction}
Inspired by Satoshi Nakamoto's idea of a direct exchange currency \cite{Nakamoto2009}, we've seen a growing interest in cryptocurrencies in recent years, with Bitcoin standing out as the first major decentralized one. Other cryptocurrencies like Ether are also becoming more popular, including stablecoins such as USDT and USDC that maintain their value pegged to the US Dollar. The foundational ethos of Bitcoin and of some of the most popular cryptocurrencies that have followed is to enable routine transactions. However, while the speculative value of these digital assets is acknowledged by many, comprehending their wider adoption and utilization in daily purchases is an emerging field of research.

A "crypto-shopper" refers to a consumer who engages in commercial transactions utilizing cryptocurrencies as a medium of exchange, often reflecting varied levels of blockchain and cryptocurrency domain knowledge and expertise. This study investigates the complexities of  "crypto-shoppers", assessing the extent of their domain knowledge and how this impacts their purchasing behaviors. By examining the levels of knowledge, expertise, and transactional behavior, we aim to provide insights into the evolving landscape of cryptocurrency adoption in commerce.

\subsection{Motivation \& Research Questions}
While the Cryptorefills' consumer reports of 2021 \cite{Cabuk2021} and 2022 \cite{Cabuk2022} have shed initial light on the interrelation between domain knowledge and crypto-shopper purchase frequency, several areas necessitate further exploration. To begin with, the reports provided a preliminary overview and were devoid of an exhaustive analysis regarding the diverse degrees of knowledge and expertise among crypto consumers. Additionally, the correlation between knowledge/expertise and purchase frequency remains largely unexplored. Although the 2022 report conducted a thorough segmentation of crypto-shoppers via its cluster analysis—encompassing a broad range of user types with varying knowledge bases, demographics, geographical influences, and perceptions—it was confined to only five distinct variables pertinent to knowledge and expertise.

Considering the diverse aspects of the crypto sector—which spans areas from technology to finance, from currency concepts to infrastructure, and from governance to emerging fields like NFTs and blockchain gaming (all represented within the 17 variables of knowledge and expertise)—there is conceivable value in conducting a segmentation that singularly emphasizes these specific dimensions of knowledge and expertise, incorporating purchase frequency behavior as an additional analytical dimension.

Using the updated dataset prepared for the upcoming Cryptorefills Labs 2023 report, which aligns knowledge levels with past reports and introduces longitudinal studies tracking the progression of knowledge, expertise, and purchase frequency over three years, this study aims to go deeper, targeting gaps in the existing literature and seeking to answer the following research questions:
\begin{enumerate}[itemsep=0.3em, parsep=0pt]
    \item [\textbf{RQ1}] Can we describe the different levels of domain knowledge and expertise among crypto-shoppers, and how do these levels manifest in their utilization of cryptocurrencies for making purchases?
    \item [\textbf{RQ2}] How do varying degrees of knowledge and expertise correlate with the purchase frequencies among crypto-shoppers? Are there recognizable patterns or trends inherent in these correlations? Is a more nuanced relationship observable between domain knowledge/expertise and purchase frequency, transcending the initial findings of the Cryptorefills Labs reports?
    \item [\textbf{RQ3}] Is it possible to identify specific clusters or groupings of crypto-shoppers based on their knowledge, expertise, and shopping frequencies, and, if identifiable, describe these clusters and how they diverge in crypto-shopping behavior?
\end{enumerate}

\section{Literature Review}
Cryptocurrencies, with Bitcoin at the forefront, have garnered substantial attention recently. Satoshi Nakamoto's initial vision, delineated in his foundational white paper \cite{Nakamoto2009}, conceptualized Bitcoin as a peer-to-peer electronic cash system, underscoring its prospective utility for routine transactions. This literature review probes into the dynamics of cryptocurrency adoption, with an emphasis on the role of knowledge in the evolution of emerging technologies and particularly investigating the knowledge level of individuals utilizing cryptocurrencies for shopping purposes.

\subsection{Adoption of Bitcoin and Other Cryptocurrencies}
As of 2023, approximately 420 million individuals hold cryptocurrencies \cite{Triplea2023}. Although predominantly perceived as speculative assets, Bitcoin and alternative cryptocurrencies {such as Ether\cite{Buterin2016}} are being utilized for routine transactions \cite{Zhang2023}. The 2022 Cryptorefills Labs Report \cite{Cabuk2022} indicates that the most important challenge consumers face while using cryptocurrencies to purchase goods and services is the absence of merchants and stores accepting them as a payment method, a concern articulated by 48.2\% of the respondents. From the perspective of merchants, a 2022 report by Deloitte \cite{Tanco2023} reflected that 75\% of CEOs projected to embrace cryptocurrency as a method of payment within the following two years. However, the data for this report were collected between December 3 and 16, 2021, amidst a phase of intensified cryptocurrency market activities and prior to a strong diminution in aggregate volume and market capitalization \cite{Coinmarketcap2023}.

As per the data published by coinmap.org in August 2023\cite{Coinmap2023}, a total of 32,139 businesses worldwide are presently acknowledging cryptocurrency. For contextual understanding, Fundera \cite{Fundera2022} disclosed in October 2022 that coinmap.org had enumerated 15,174 businesses that were accepting cryptocurrency, representing a substantial augmentation of 111.8\% within a ten-month timeframe.

\subsection{Knowledge and Technological Adoption}
The adoption of new technologies has long been an important domain of research within different fields of business and innovation. Everett Rogers' text, "Diffusion of Innovations"\cite{Rogers2003} (first published in 1962), has played an important role in framing our comprehension of technological adoption mechanisms. Rogers accentuates knowledge as a paramount element in the propagation of innovations, outlining the phases that individuals go through in assimilating new technologies and emphasizing the vital roles of knowledge assimilation, communicative conduits, and societal structures in this progression.

Fred Davis’ Technology Acceptance Model \cite{Davis1989} proposes a paradigm that builds upon the predicates of technological assimilation. While the model doesn’t exclusively concentrate on knowledge, it accentuates the relevance of perceived utility and perceived user-friendliness, both intimately connected to an individual’s awareness and comprehension of technology. This acknowledgment amplifies the inference of knowledge being pivotal in shaping adoption decisions.

Lundvall’s investigation of the role of knowledge in the innovation process \cite{Lundvall2007} explores the interconnection between knowledge and innovation, emphasizing distinct kinds of knowledge—implicit and explicit—and their relevance throughout various junctures of the innovation trajectory, including technological assimilation. This analysis emphasizes the instrumental role of learning, knowledge dissemination, and absorptive capacity in aiding the assimilation of emergent technologies.

Neil Pollock and Robin Williams’ research \cite{Pollock2010} redirects the focus towards digital innovations and knowledge allocation within organizational frameworks. Their research describes the collaborative contribution of diverse actors in their knowledge and expertise, mutually defining the adoption and execution of innovative technologies. The authors outline the intricate dynamics among knowledge, labor segmentation, and technological assimilation. A holistic review by Wisdom et al. \cite{Wisdom2013} structures various theoretical stances on innovation assimilation, describing how knowledge is integral in decision-making paradigms pertinent to technological assimilation.

Knowledge holds an important role in fostering the adoption of groundbreaking technologies, and this includes the case of using blockchain and cryptocurrencies for commerce and shopping. The acquisition and propagation of knowledge modulate individuals’ resolutions to adopt and incorporate such innovations. The symbiotic relationship between education, comprehension, expertise, and organizational dynamics emphasizes the role of knowledge in maximizing the utility and advantages of novel technological breakthroughs.

\subsection{Domain-Specific Knowledge and Expertise of Crypto-Shoppers}
Literature investigating the knowledge and expertise of cryptocurrency users, particularly those utilizing cryptocurrencies for shopping, remains scarce. A study executed by Alqaryouti et al. \cite{Alqaryouti2020} employed interviews with three cryptocurrency users, disclosing that the majority exhibited a comprehensive grasp of cryptocurrency, including intricate technical aspects such as mining and market capitalization. The participants predominantly perceived cryptocurrency as a medium of investment and a currency, praising its advantages, including decentralization, security, and cost efficiency, while also acknowledging the global acceptance limitations as a notable disadvantage. Despite the restricted sample size, this study focused more on the general cryptocurrency user base instead of concentrating explicitly on those utilizing cryptocurrencies for shopping.

A more relevant investigation, the 2021 report by Cryptorefills Labs \cite{Cabuk2021}, probed into the particular knowledge and expertise terrain of crypto-shoppers, outlining five knowledge dimensions and four expertise parameters pertinent to this demographic. The report illustrated that a substantial portion of crypto-shoppers possess familiarity with foundational blockchain and cryptocurrency principles, with the majority being acquainted with Bitcoin and its functionalities, while more advanced technical comprehension, like the understanding of the Lightning Network, is predominantly held by a more specialized subset. It established a correlational relationship between crypto-consumers’ expertise in cryptocurrency trading and their shopping frequencies, revealing that approximately three-quarters of the experts effectuate crypto-purchases weekly.

\renewcommand{\arraystretch}{0.88}
\setlength{\tabcolsep}{.25em}
\begin{table}[!b]
\caption{Survey Questions}
\label{tab:questions}
\begin{tabular}{@{}llp{2.85in}@{}}
\toprule
\multicolumn{2}{l}{\textit{\textbf{Task}}} & \textit{Please rate the extent to which you agree/disagree with the following statements using 5-point Likert scale.} \\ \toprule
\textbf{Q. \#} & \textbf{Q. ID} & \textbf{Question/Statement}                                                 \\ \toprule
3                  & K.1               & I have a general understanding of what a blockchain is   and how it works   \\
4                  & K.2               & I have a general understanding of what bitcoin is and   how it works        \\
5                  & K.3               & I know how to obtain bitcoin                                                \\
6                  & K.4               & I know how to obtain some other cryptocurrencies that are not Bitcoin       \\
7                  & K.5               & I have a general understanding of what the Lightning Network is             \\
8                  & E.1               & I am a blockchain technology expert                                         \\
9                  & E.2               & I am a bitcoin expert                                                       \\
10                 & E.3               & I am a cryptocurrency expert                                                \\
11                 & E.4               & I am a cryptocurrency investment and/or trading expert                      \\
12                 & K.6               & I know how to use a Layer-2 Blockchain Network   (Polygon, Avalanche, etc.) \\
13                 & K.7               & I have a general understanding of what an NFT is                            \\
14                 & E.5               & I am an NFT expert                                                        \\
15                 & K.8               & I play blockchain-based games                                               \\
16                 & K.9               & I have a general understanding of what a DAO is                             \\
17                 & K.10               & I participate in one or more DAOs                                           \\
18                 & K.11              & I know how to use DeFi services                                             \\
19                 & K.12              & I am a DeFi expert                                                          \\ \bottomrule
\end{tabular}
\end{table}

\begin{table*}[!b]
\caption{Knowledge/Expertise and Purchase Frequency (Bottom Row) Descriptive Statistics} \centering
\label{tab:descriptive}
\begin{tabular}{@{}lcccccccccc@{}}
\toprule
\textbf{Question (Shortened)} &
  \textbf{count} &
  \textbf{mean} &
  \textbf{std. dev.} &
  \textbf{min} &
  \textbf{25\%} &
  \textbf{median} &
  \textbf{75\%} &
  \textbf{max} &
  \textbf{mode} &
  \textbf{mode count} \\ \toprule
Understanding of   blockchain                    & 516 & 3.69 & 1.41 & 1 & 3 & 4 & 5 & 5 & 5 & 204 \\
Understanding of bitcoin                         & 516          & 3.82          & 1.37 & 1 & 3 & 4 & 5 & 5 & 5 & 229 \\
Knowledge on obtaining   bitcoin                 & 516          & 3.82          & 1.44 & 1 & 3 & 4 & 5 & 5 & 5 & 251 \\
Knowledge on obtaining   other cryptocurrencies  & 516          & 3.83          & 1.41 & 1 & 3 & 4 & 5 & 5 & 5 & 249 \\
Understanding of the   Lightning Network         & 516          & 3.23          & 1.38 & 1 & 2 & 3 & 4 & 5 & 5 & 124 \\
Expertise in blockchain   technology             & 516          & 2.76          & 1.25 & 1 & 2 & 3 & 4 & 5 & 3 & 144 \\
Expertise in bitcoin                             & 516          & 2.91          & 1.29 & 1 & 2 & 3 & 4 & 5 & 3 & 141 \\
Expertise in   cryptocurrency                    & 516          & 3.07          & 1.31 & 1 & 2 & 3 & 4 & 5 & 3 & 142 \\
Expertise in   cryptocurrency investment/trading & 516          & 3.04          & 1.31 & 1 & 2 & 3 & 4 & 5 & 4 & 129 \\
Knowledge on using  layer-2 blockchain network  & 516          & 3.12          & 1.45 & 1 & 2 & 3 & 4 & 5 & 5 & 126 \\
Understanding of NFTs                            & 516          & 3.45          & 1.37 & 1 & 2 & 4 & 5 & 5 & 5 & 149 \\
Expertise in NFTs                                & 516          & 2.76          & 1.3  & 1 & 2 & 3 & 4 & 5 & 2 & 130 \\
Playing blockchain-based  games                 & 516          & 2.83          & 1.43 & 1 & 2 & 3 & 4 & 5 & 1 & 124 \\
Understanding of DAOs                            & 516          & 2.86          & 1.4  & 1 & 2 & 3 & 4 & 5 & 1 & 117 \\
Participation in DAOs                            & 516          & 2.73          & 1.4  & 1 & 1 & 3 & 4 & 5 & 1 & 134 \\
Knowledge on using DeFi  services               & 516          & 3.09          & 1.45 & 1 & 2 & 3 & 4 & 5 & 5 & 118 \\
Expertise in DeFi                                & 516          & 2.67          & 1.32 & 1 & 2 & 3 & 4 & 5 & 2 & 129 \\
Purchase frequency                               & 445          & 3.26          & 1.05 & 1 & 3 & 3 & 4 & 5 & 3 & 171 \\ \bottomrule
\end{tabular}
\end{table*}

The 2022 iteration of the study \cite{Cabuk2022} introduced six additional determinants to evaluate the knowledge and expertise of crypto-shoppers, incorporating areas such as non-fungible token (NFT) expertise and a comprehensive understanding of decentralized autonomous organizations (DAO) \cite{Chistiakov2020}. The insights derived indicate that, as of 2022, the knowledge of blockchain and cryptocurrencies continues to play a crucial role in shaping the shopping habits of crypto-consumers. A notable segment maintains a foundational understanding of blockchain principles and cryptocurrencies, with approximately 70\% recognizing Bitcoin, its acquisition methodologies, and its sustained market dominance. A significant correlation was identified between shoppers’ cryptocurrency expertise and their purchasing frequency, with the preponderance of experts engaging in crypto-based transactions weekly emphasizing the perpetual importance of well-versed crypto-consumers.

For the first time, the 2022 study employed a K-prototypes clustering method on 29 disparate variables from the survey data to categorize the varied crypto user base, introducing a nuanced cluster analysis of Crypto-shoppers. Amongst these variables, five were explicitly associated with knowledge and expertise, with other variables including demographics, geographic locations, and notably, constructs associated with the Technology Acceptance Model 2 (TAM2) such as "Perceived Ease of Use", "Perceived Usefulness", and "Social Influence". The discernment of seven distinct clusters in the 2022 report indicates the diversity and complexity within the crypto user demographic, suggesting a spectrum of user categories, each its unique combinations of knowledge, demographic traits, geographical influences, and perceptions delineated by TAM2 constructs and therefore transcending simplistic categorizations such as financially excluded users from emerging markets versus technologically savvy users from advanced markets.

\section{Methodology} \label{sec:method}
\subsection{Survey Design}
The study extracts its primary dataset from the 2023 Cryptorefills Labs report, which is under preparation at the time of writing. The survey, an integral component of the report, was made available to Cryptorefills customers from June 1st, 2023, to September 9th, 2023, collecting data only from customers who have successfully executed at least one purchase, confirming their authentic crypto-shopper status and ensuring alignment with the scope of our posed research questions.

The survey is structured into 145 multiple-choice questions segmented to facilitate a comprehensive exploration of the respondent's behaviors, demographics, perceptions, and experiences and is based on theoretical frameworks like TAM2 (Technology Acceptance Model 2) as well as managerial insights from the Cryptorefills team. The questions include but are not limited to demographics, cryptocurrency familiarity, and multiple constructs from TAM2, such as Perceived Ease of Use (PEO) and Perceived Usefulness (PU), ensuring a holistic understanding of cryptocurrency shopping adoption. A special emphasis was placed on 17 questions (see Table \ref{tab:questions} for the complete list), which focus on knowledge and expertise, utilizing a 5-point Likert scale to gauge respondents’ levels of agreement or disagreement with various statements. In the adopted scale, point '1' corresponds to "mostly disagreeing", '2' corresponds to "somewhat disagreeing", '3' corresponds to "feeling neutral", '4' corresponds to "somewhat agreeing", and '5' corresponds to "mostly agreeing" with the given statement or question.

\subsection{Data Collection}
The survey data was collected utilizing the Computer Assisted Web Interviewing (CAWI) technique through the Cryptorefills web platform. The platform enabled the acquisition of insights specifically from authentic users with a recorded history of cryptocurrency transactions, consolidating the study’s reliability. Google Forms was the chosen medium to host the survey, ensuring a seamless experience for the participants.

Cryptorefills implemented its loyalty program to encourage participation, allocating 300 points to each respondent, equivalent to just below \$3 USD. It is important to highlight that this allocation wasn't substantial enough to claim rewards on Cryptorefills in an effort to maximize responses’ authenticity. A multi-channel promotional strategy was also integrated, utilizing email, Twitter promotions, and website placements to enhance awareness and encourage customer participation.

\subsection{Sample Size}
The elimination of duplicate entries refined the dataset to 516 unique responses collected between June 3rd and September 3rd, 2023. This specialized dataset provides a representative snapshot of the transactions conducted by 7,535 distinct customers who used the web platform for various reasons during the designated timeframe, with a calculated margin of error at approximately $\pm$4.16\% with a 95\% confidence interval (CI), allowing for reliable data extrapolations and detailed analysis.

\subsection{Data Analysis}
The analytical phase initiated with rigorous preprocessing of the data to validate its consistency, reliability, and validity. A Cronbach's alpha test yielded a score of 0.967, confirming a very high level of internal consistency within the dataset. Descriptive statistics were employed to provide a nuanced overview of the knowledge and expertise levels of crypto-shoppers. The analytical approach also consisted of undertaking multiple regression analyses to investigate the relationships and the K-means clustering technique to identify distinct segments of crypto-shoppers based on shared attributes and behaviors. The findings were articulated and clarified using a plethora of visual tools, including, but not limited to, bar charts, scatter plots, and radar plots, ensuring enhanced clarity and facilitating the nuanced interpretation of the data.

\section{Results}
\subsection{Descriptive Statistics}
The analytical overview provided by the descriptive statistics unveils valuable insights into the respondents' varied knowledge and expertise levels in the domains of blockchain and cryptocurrency. The utilization of the Likert scale responses, assessed from a spectrum of 1 to 5, allows 
 to assess central tendencies and dispersion.

\begin{figure}[!t]
 \centering
 \includegraphics[scale=0.366]{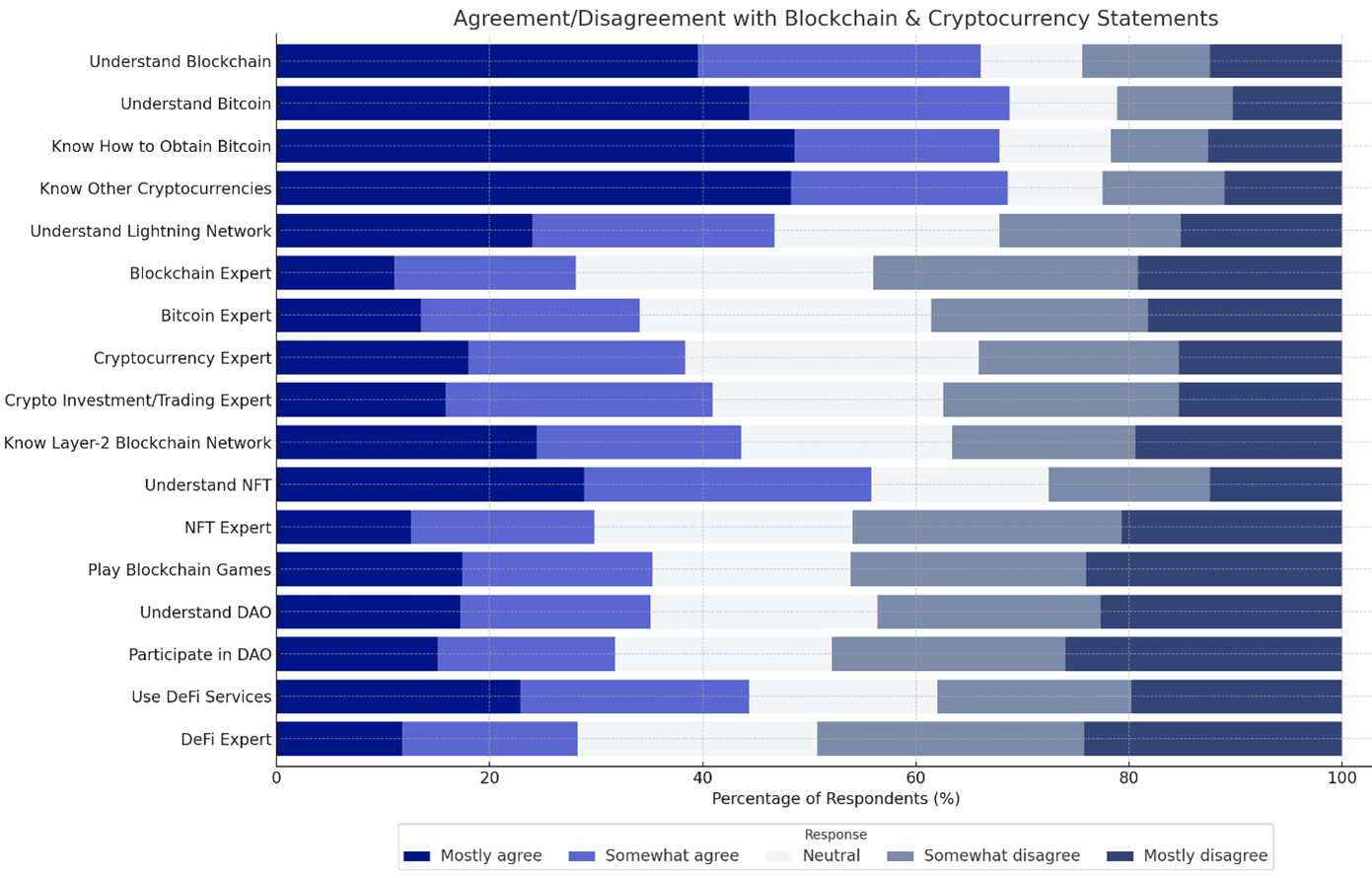}
 \caption{Stacked distribution of responses on blockchain \& cryptocurrency expertise.}
 \label{fig:fig2}
\end{figure}

\begin{figure}[!b]
 \centering
 \includegraphics[scale=0.515]{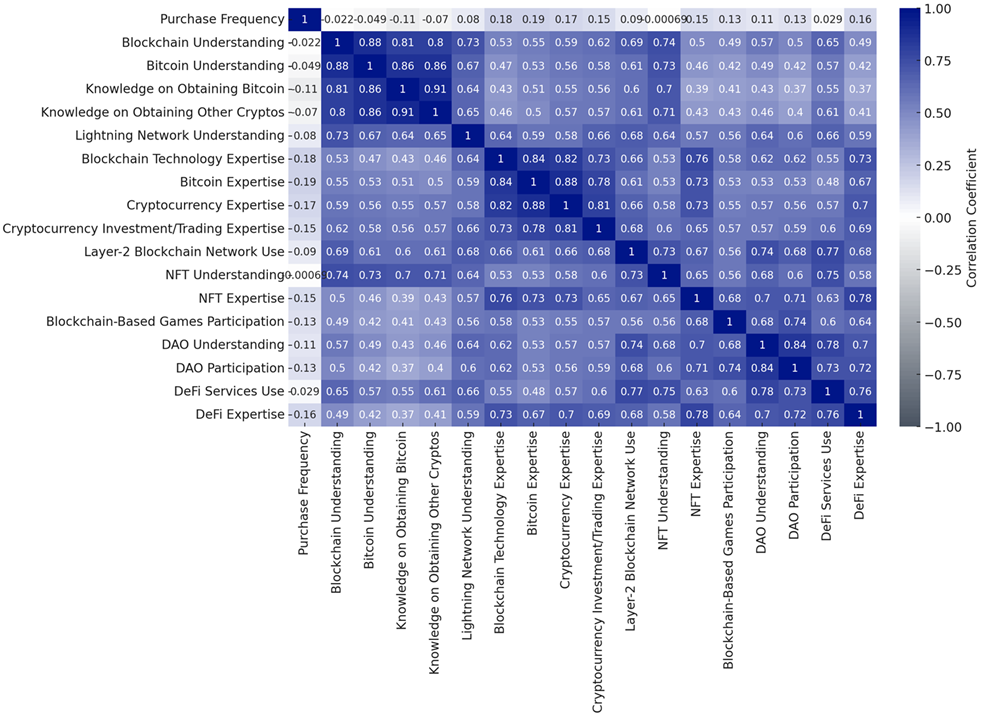}
 \caption{Correlation matrix and heatmap of the 17 knowledge variables and the purchase frequency variable.}
 \label{fig:fig4}
\end{figure}

\begin{figure*}[!tb]
 \centering
 \includegraphics[scale=.9]{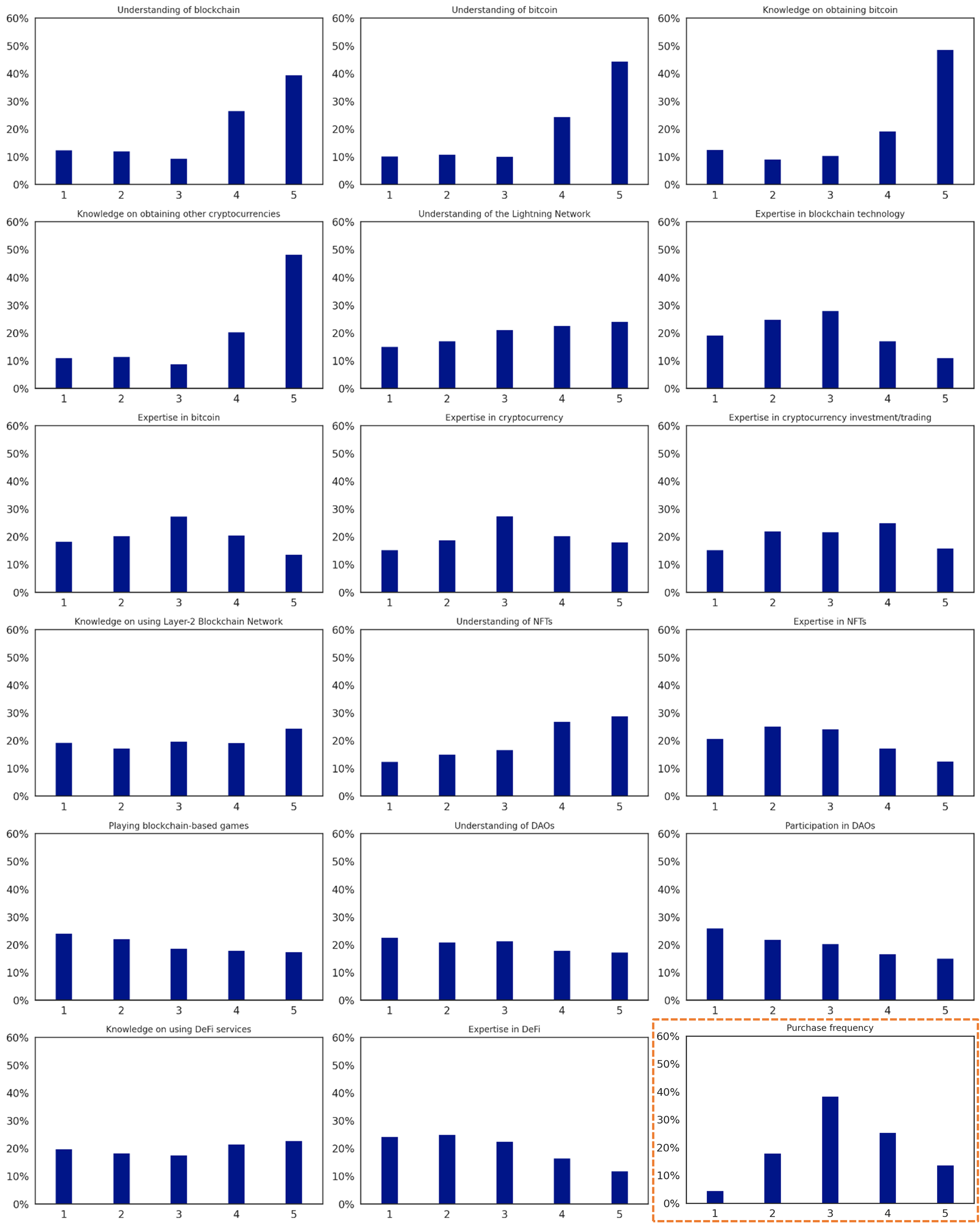}
 \caption{Distribution of knowledge/expertise and purchase frequency (last chart with an orange border) responses based on a 5-point Likert scale.}
 \label{fig:fig1}
\end{figure*}

It’s clear that a considerable proportion of respondents possess general knowledge of the foundational elements of cryptocurrency, evidenced by elevated average ratings concerning their understanding of both blockchain and bitcoin, noted at 3.69 and 3.82, respectively. Both areas also see the highest rating of 5 as the most frequent response.
Unlike foundational knowledge of Bitcoin and other cryptocurrencies, when participants were asked about their self-perceived expertise (as opposed to general understanding) in Bitcoin or other cryptocurrencies, the average response hovered around a score of 3.

Going deeper into more specialized areas of the crypto world, the level of understanding begins to vary and decrease. For instance, the understanding of the Lightning Network averages 3.23, while the professed expertise in blockchain technology dips to an average of 2.76. 

\begin{figure*}[!tb]
 \centering
 \includegraphics[scale=0.9]{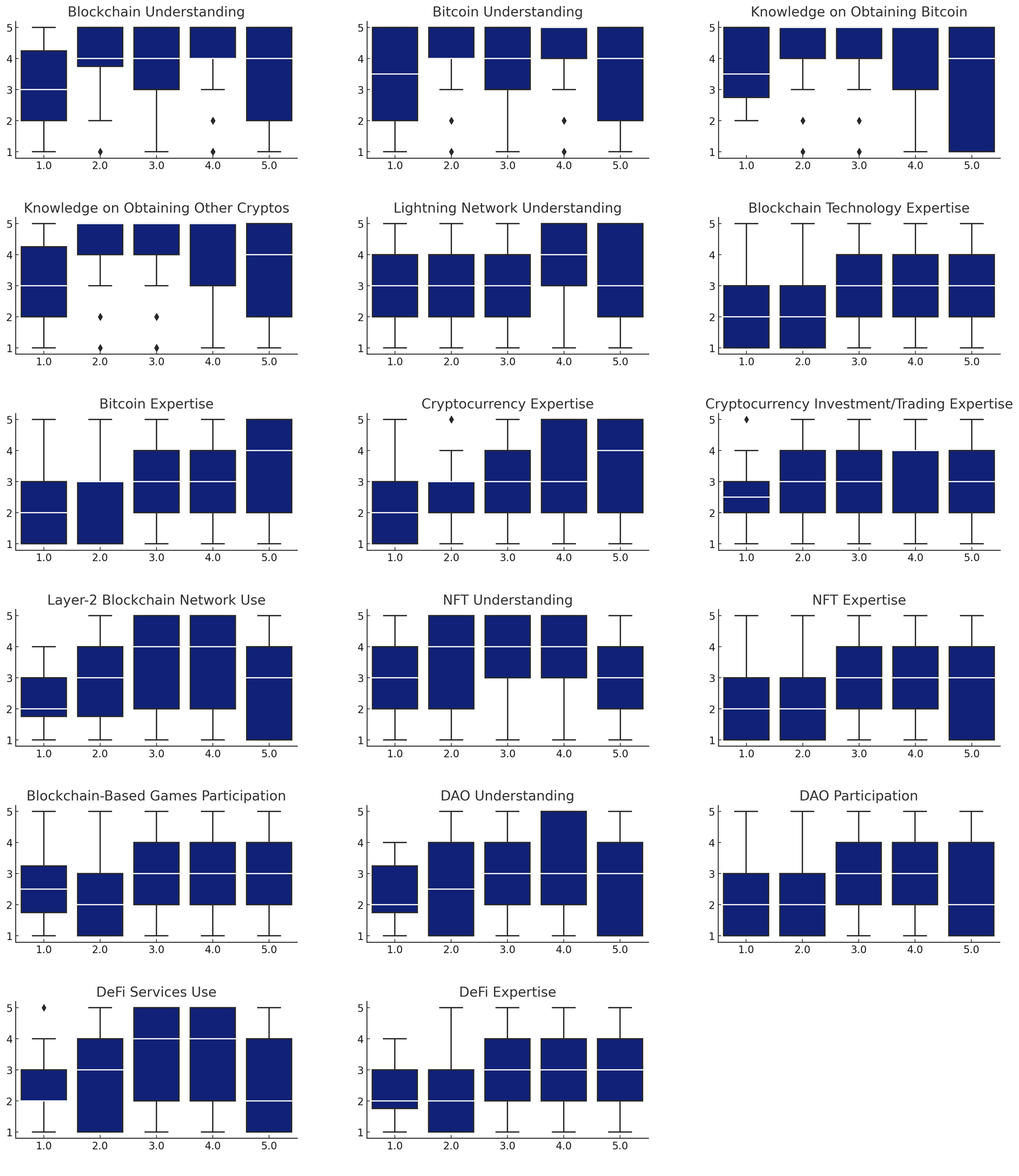}
 \caption{Box plot analysis of the purchase frequency concerning different aspects of knowledge and expertise in blockchain and cryptocurrency matters.}
 \label{fig:fig3}
\end{figure*}

NFTs, or Non-Fungible Tokens, present another interesting case. While the average self-reported understanding is 3.45, the expertise in NFTs is only 2.76, indicating that while many are aware of what NFTs are, fewer feel confident in their deep knowledge about them. Similarly, for DAOs (Decentralized Autonomous Organizations), the understanding is moderately rated at 2.86, but it's interesting to see that the lowest score is the most common response. A significant subset of the respondents is, therefore, unfamiliar with DAOs.

Regarding crypto’s intersection with gaming and DeFi services, the average knowledge levels are relatively balanced, resonating at 2.83 and 3.09, respectively. This implies a diverse knowledge landscape, ranging from fundamental to advanced, and from theoretical to applied, indicative of the growing diversification and evolution of the crypto ecosystem. More details can be found in Table \ref{tab:descriptive}.

\subsection{Visual Analysis}
The descriptive statistics are further complemented by a series of charts, visually representing the knowledge distribution among the respondents. Figures \ref{fig:fig2} and \ref{fig:fig1} provide a distribution of responses for self-claimed expertise on blockchain and cryptocurrencies and the purchase frequency, constituting a basis for further correlation analysis. The notable skewness towards higher levels of understanding for foundational topics juxtaposed against the broader distribution for niche areas sheds light on the knowledge disparities within the crypto community. Figure \ref{fig:fig3}, on the other hand, visualizes the purchase frequency per different factors of knowledge and expertise in various blockchain applications.

The distribution patterns for engagement with specialized cryptocurrency applications or governance models describe a contrasting trend, revealing a dominance of lower-end peaks indicative of potential knowledge-application gaps in these advanced areas. The frequency distribution for cryptocurrency purchase frequency depicts a central tendency, highlighting a balanced, moderate utilization of cryptocurrency for transactions.

\subsection{Correlation and Regression Analysis}
To explore the multidimensional relationships between respondents' knowledge and their purchasing behavior, correlation and regression analyses were implemented. The analysis revealed a series of negative correlations between purchase frequency and all knowledge variables, with "Knowledge on Obtaining Bitcoin" yielding the strongest negative correlation at -0.11. This highlights a nuanced relationship where increments in knowledge or expertise in specific areas are subtly correlated with a decrement in purchase frequency. The complete correlation matrix is given in Figure \ref{fig:fig4}.

The regression model presented in detail in Table \ref{tab:reg} provides a broader perspective, explaining approximately 11.63\% of the variance in the dependent variable. Despite its statistical significance, evidenced by the substantially small p-value (p<0.001), the model exhibits a limited explanatory power, reflected by the R-squared values. The individual predictors within the model bring forth "Knowledge on Obtaining Bitcoin" as a singularly significant variable, indicative of its nuanced impact on purchase frequency.

\begin{table*}[!b]
\caption{Regression Analysis: Can Purchase Frequency Be Explained by Knowledge?} \centering
\label{tab:reg}
\begin{tabular}{@{}lccccc@{}}
\toprule
\multicolumn{6}{c}{\textbf{Model Summary}}                                                               \\ \midrule
Dependent Variable (Question) &
  \multicolumn{5}{c}{I purchase products   and/or services paying with bitcoin or other crypto} \\
R-squared                                   & \multicolumn{5}{c}{0.116}                                  \\
Adjusted R-squared                          & \multicolumn{5}{c}{0.081}                                  \\
F-statistic                                 & \multicolumn{5}{c}{3.306}                                  \\
Prob (F-statistic)                          & \multicolumn{5}{c}{0.000011}                               \\
Log-Likelihood                              & \multicolumn{5}{c}{-624.11}                                \\
AIC (Akaike Information   Criterion)        & \multicolumn{5}{c}{1284.2}                                 \\
BIC (Bayesian Information   Criterion)      & \multicolumn{5}{c}{1358.0}                                 \\ \midrule
\multicolumn{6}{c}{\textbf{Calculated Coefficients}}                                                     \\ \midrule
\textit{\textbf{Question (Shortened)}} &
  \textit{\textbf{Coefficient}} &
  \textit{\textbf{Std. Error}} &
  \textit{\textbf{t-value}} &
  \textit{\textbf{p-value}} &
  \textit{\textbf{95\% Confid. Interv. {[}0.025, 0.975{]}}} \\ \midrule
const                                       & 3.2069  & 0.1573 & 20.381 & 0     & {[}2.8976, 3.5161{]}   \\
Blockchain Understanding                    & -0.0255 & 0.0845 & -0.301 & 0.763 & {[}-0.1916, 0.1407{]}  \\
Bitcoin Understanding                       & 0.0258  & 0.0965 & 0.268  & 0.789 & {[}-0.1638, 0.2154{]}  \\
Knowledge on Obtaining   Bitcoin            & -0.2981 & 0.091  & -3.277 & 0.001 & {[}-0.4769, -0.1193{]} \\
Knowledge on Obtaining   Other Cryptos      & 0.0418  & 0.0921 & 0.454  & 0.65  & {[}-0.1392, 0.2229{]}  \\
Lightning Network   Understanding           & 0.0886  & 0.0619 & 1.431  & 0.153 & {[}-0.0331, 0.2103{]}  \\
Blockchain Technology   Expertise           & -0.0368 & 0.0855 & -0.43  & 0.667 & {[}-0.2048, 0.1313{]}  \\
Bitcoin Expertise                           & 0.1192  & 0.0926 & 1.288  & 0.198 & {[}-0.0627, 0.3012{]}  \\
Cryptocurrency Expertise                    & 0.1263  & 0.0918 & 1.376  & 0.17  & {[}-0.0541, 0.3066{]}  \\
Cryptocurrency   Investment/Trading Expertise &
  0.0136 &
  0.0733 &
  0.185 &
  0.853 &
  {[}-0.1304, 0.1576{]} \\
Layer-2 Blockchain   Network Use            & 0.0751  & 0.064  & 1.174  & 0.241 & {[}-0.0506, 0.2009{]}  \\
NFT Understanding                           & -0.0061 & 0.0708 & -0.086 & 0.932 & {[}-0.1453, 0.1331{]}  \\
NFT Expertise                               & -0.0711 & 0.0789 & -0.902 & 0.368 & {[}-0.2262, 0.0839{]}  \\
Blockchain-Based Games   Participation      & 0.0622  & 0.0538 & 1.156  & 0.248 & {[}-0.0435, 0.1680{]}  \\
DAO Understanding                           & 0.0329  & 0.0738 & 0.446  & 0.656 & {[}-0.1121, 0.1779{]}  \\
DAO Participation                           & -0.0085 & 0.071  & -0.119 & 0.905 & {[}-0.1480, 0.1310{]}  \\
DeFi Services Use                           & -0.1108 & 0.0768 & -1.443 & 0.15  & {[}-0.2618, 0.0401{]}  \\ \midrule
\multicolumn{6}{c}{\textbf{Diagnostics}}                                                                 \\ \midrule
Durbin-Watson Test (for   autocorrelation)  & \multicolumn{5}{c}{1.967}                                  \\
Omnibus Test (for normality of   residuals) & \multicolumn{5}{c}{5.315}                                  \\
Prob. (Omnibus)                               & \multicolumn{5}{c}{0.070}                                  \\
Skew (of residuals)                         & \multicolumn{5}{c}{-0.182}                                 \\
Kurtosis (of residuals)                     & \multicolumn{5}{c}{-0.339}                                 \\
Jarque-Bera (JB) Test                       & \multicolumn{5}{c}{4.694}                                  \\
Prob. (JB)                                    & \multicolumn{5}{c}{0.096}                                  \\
Condition Number                            & \multicolumn{5}{c}{47.2}                                   \\ \bottomrule
\end{tabular}
\end{table*}

\subsection{Cluster Analysis}
Utilizing the K-means cluster analytical approach, the variation in blockchain knowledge among respondents was examined. The elbow method, as visualized in Figure \ref{fig:fig5}, suggested potential optimal cluster solutions either as three or four. Post-analysis, the three-cluster paradigm was favored for its superior coherence and distinction. Results presented in Table \ref{tab:cluster} confirm that the three-cluster structure presents a sharper demarcation, with a pronounced clarity in the low-knowledge cohort, as depicted in Figure \ref{fig:fig6}. Essential cluster metrics endorse the three-cluster methodology's superiority, characterized by a higher silhouette score, a lower Davies Bouldin index, and a higher Calinski Harabasz score. The clusters can be detailed as follows:
\begin{figure}[!tb]
 \centering
 \includegraphics[scale=.51]{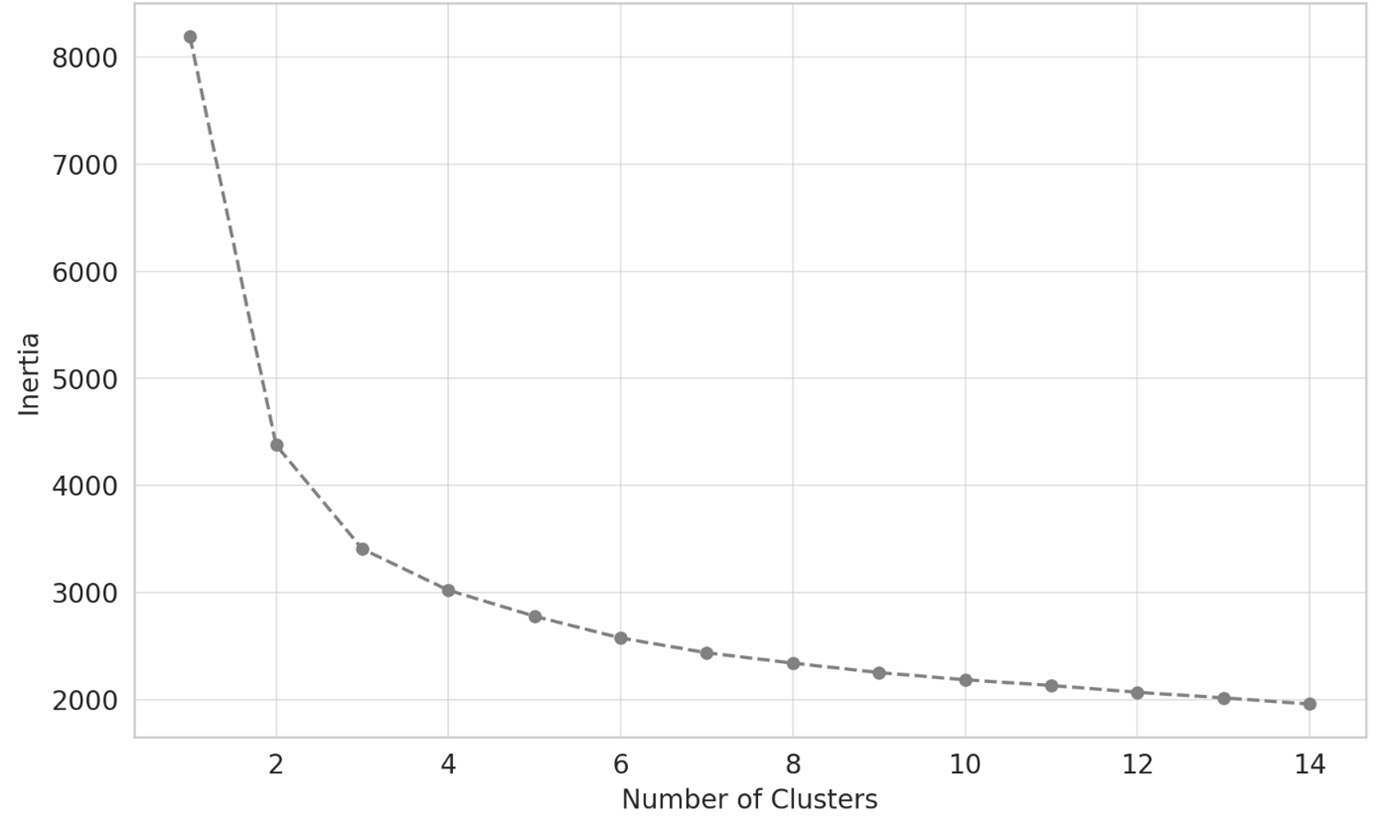}
 \caption{Elbow method showing the optimal number of consumer clusters.}
 \label{fig:fig5}
\end{figure}
\begin{itemize} [itemsep=0.3em, parsep=0pt]
    \item \textbf{Cluster 0: Low Knowledge Group} (28.09\% of the population), this cohort displays minimal knowledge concerning cryptocurrency and associated paradigms. Foundational constructs such as blockchain and bitcoin fluctuate between 1.8 and 2.1, implying elementary understanding. As the subject matter intensifies in specialization, their knowledge remains modest. Their self-rated expertise in particular areas, like blockchain or bitcoin expertise, predominantly falls below 2, signifying constrained assurance in their capabilities. Intriguingly, despite their shallow depth of knowledge, their crypto market engagement is substantial, with a purchasing frequency approximation of 3.27.
    \item \textbf{Cluster 1: Expert Group} (35.06\% of the population), members of this segment consistently manifest profound knowledge in cryptocurrency-related topics. Foundational pillars such as blockchain and bitcoin approach ratings of approximately 4.8, indicating a robust foundation. Their expertise also spans specialized niches like DeFi and NFTs, registering ratings beyond 4. Self-perceptions regarding expertise in diverse domains typically surpass 4, suggesting heightened confidence potentially stemming from vast crypto-domain immersion. Their participation levels, notably around 4, in specific blockchain and crypto applications like DAO involvement, blockchain gaming, and DeFi utilization, are high. Their purchasing frequency approximates 3.51, marginally surpassing the Low Knowledge Group.
    \item \textbf{Cluster 2: Moderate Knowledge Group} (36.85\% of the population), these individuals exhibit a decent grasp of cryptocurrency basics, with metrics around 4.2 for both blockchain and bitcoin. However, their understanding wanes slightly for advanced topics, though they maintain an above-average stance. Their self-assessment in niche sectors hints at moderate expertise. For example, while familiar with the concept of NFTs, their expertise self-rating rests around 2.33. Their purchasing cadence is approximately 2.98, subtly inferior to the other clusters.
\end{itemize}

\begin{table}[!tb]
\caption{K-Means Cluster Analysis} \centering
\label{tab:cluster}
\begin{tabular}{@{}p{1.95in}ccc@{}}
\toprule
\multicolumn{4}{c}{\textbf{Cluster Means}}                                              \\ \midrule
\textit{\textbf{Question (Shortened)}}      & \textit{\textbf{Cluster 0}}   & \textit{\textbf{Cluster 1}} & \textit{\textbf{Cluster 2}} \\ \midrule
Blockchain Understanding                    & 1.888  & 4.8049         & 4.1667        \\
Bitcoin Understanding                       & 2.128  & 4.7744         & 4.3718        \\
Knowledge on Obtaining Bitcoin              & 2.136  & 4.7256         & 4.5321        \\
Knowledge on Obtain. other Crypto        & 2.104  & 4.7683         & 4.4872        \\
Lightning Network Understanding             & 1.848  & 4.4085         & 3.3205        \\
Blockchain Technology Expertise             & 1.68   & 3.9451         & 2.4423        \\
Bitcoin Expertise                           & 1.712  & 4.0427         & 2.7756        \\
Cryptocurrency Expertise                    & 1.752  & 4.2744        & 2.9744        \\
Cryptocur. Invest./Trade Expertise & 1.68   & 4.2195         & 2.968        \\
Layer-2 Blockchain Network Use              & 1.568  & 4.4085         & 3.1346        \\
NFT Understanding                           & 1.872  & 4.5488         & 3.75           \\
NFT Expertise                               & 1.688  & 4.0122          & 2.3333        \\
Blockchain-Based Games Particip.        & 1.784  & 4.1342         & 2.3526        \\
DAO Understanding                           & 1.72   & 4.2012         & 2.5192        \\
DAO Participation                           & 1.672  & 4.0793         & 2.2628        \\
DeFi Services Use                           & 1.64   & 4.3537         & 3.0897        \\
DeFi Expertise                              & 1.664  & 3.9634         & 2.2115        \\
Purchasing with Crypto                      & 3.272  & 3.5122          & 2.9808        \\ \midrule
\multicolumn{4}{c}{\textbf{ANOVA for Cluster Differences}}                              \\ \midrule
\textit{\textbf{Question (Shortened)}}      & \textit{\textbf{F-statistic}} & \multicolumn{2}{c}{\textit{\textbf{p-value}}}             \\ \midrule
Blockchain Understanding                    & 321.02 & \multicolumn{2}{c}{4.70 x $10^{-69}$} \\
Bitcoin Understanding                       & 332.92 & \multicolumn{2}{c}{1.85 x $10^{-70}$} \\
Knowledge on Obtaining Bitcoin              & 333.03 & \multicolumn{2}{c}{1.80 x $10^{-70}$} \\
Knowledge on Obtain. other Crypto        & 434.61 & \multicolumn{2}{c}{3.79 x $10^{-81}$} \\
Lightning Network Understanding             & 170.35 & \multicolumn{2}{c}{4.96 x $10^{-47}$} \\
Blockchain Technology Expertise             & 141.1  & \multicolumn{2}{c}{1.91 x $10^{-41}$} \\
Bitcoin Expertise                           & 99.77  & \multicolumn{2}{c}{2.17 x $10^{-32}$} \\
Cryptocurrency Expertise                    & 144.08 & \multicolumn{2}{c}{4.84 x $10^{-42}$} \\
Cryptocur. Invest./Trade Expertise & 159.83                        & \multicolumn{2}{c}{4.33 x $10^{-45}$}                          \\
Layer-2 Blockchain Network Use              & 201.43 & \multicolumn{2}{c}{2.20 x $10^{-52}$} \\
NFT Understanding                           & 210.18 & \multicolumn{2}{c}{8.49 x $10^{-54}$} \\
NFT Expertise                               & 135.25 & \multicolumn{2}{c}{2.96 x $10^{-40}$} \\
Blockchain-Based Games Particip.        & 116.96 & \multicolumn{2}{c}{2.40 x $10^{-36}$} \\
DAO Understanding                           & 152.45 & \multicolumn{2}{c}{1.10 x $10^{-43}$} \\
DAO Participation                           & 137.69 & \multicolumn{2}{c}{9.41 x $10^{-41}$} \\
DeFi Services Use                           & 151.32 & \multicolumn{2}{c}{1.83 x $10^{-43}$} \\
DeFi Expertise                              & 116.65 & \multicolumn{2}{c}{2.81 x $10^{-36}$} \\
Purchasing with Crypto                      & 14.3   & \multicolumn{2}{c}{1.34 x $10^{-06}$} \\ \midrule
\multicolumn{4}{c}{\textbf{Cluster Analysis Summary}}                                   \\ \midrule
Within-Cluster Sum of Sq. (WCSS)   & \multicolumn{3}{c}{938.8622}             \\
Between-Cluster Sum of Sq. (BCSS)       & \multicolumn{3}{c}{2507.1378}            \\
Total Variance                              & \multicolumn{3}{c}{4446.0}                \\
Proportion Explained                        & \multicolumn{3}{c}{0.5639}      \\   \bottomrule     
\end{tabular}
\end{table}

\section{Discussion}
This section integrates the research findings with prevailing academic discourse. We aim not only to provide answers to the research questions posited at the outset but also to identify the broader business implications arising from these results, ensuring their relevance and applicability in real-world context.

\subsection{Findings on RQ1: Levels of Knowledge and Expertise among Crypto-shoppers}
In addressing this research question centered on the delineation of knowledge and expertise among crypto-shoppers, the current study unveils a picture that both supports and, in some instances, challenges prevailing scholarly narratives.

Alqaryouti et al. \cite{Alqaryouti2020} underscored a marked level of cryptocurrency knowledge among their participants, an inference that finds resonance in the present dataset. Median scores, reflective of knowledge surrounding both blockchain and Bitcoin, situate at 4 within a 1 to 5 evaluative spectrum, evincing a noteworthy degree of knowledge within the crypto-shopper cohort. Such elevated median values harmonize with the conclusions drawn from the Cryptorefills Labs studies conducted in 2021 and 2022 \cite{Cabuk2021,Cabuk2022}, wherein a preponderance of crypto consumers displayed an affinity for core blockchain and cryptocurrency tenets. Nonetheless, a distinction emerges upon examining expertise. Median scores associated with "Expertise in Blockchain technology", "Expertise in Bitcoin", and "Expertise in Cryptocurrency" converge at 3, suggesting that while many crypto-shoppers possess foundational knowledge, a categorization as definitive experts remains elusive for a substantial proportion.

Investigating deeper into the metrics for "Understanding of blockchain" and "Understanding of bitcoin", a clear subset of respondents, constituting approximately a third, rate their knowledge in the low tiers (scaled at 1 or 2). This pattern suggests that a sizable faction of crypto-shoppers engage in cryptocurrency transactions without a profound grasp of the core technology. Such a trend, viewed through the adoption perspective, perhaps signals a transition from early blockchain enthusiasts and connoisseurs to individuals for whom the technology addresses tangible needs without the impetus for in-depth knowledge.

Understanding payment methods considered pivotal for blockchain payment scalability, notably the Lightning Network and Layer-2 Blockchain, presents a median index of 3. Such scores, albeit moderate, trail behind the general blockchain knowledge benchmarks, indicative of the specialized nature of these themes, and are mainly understood by those with higher expertise by those driven by pragmatic utility needs.

Interestingly, the median metric for "Understanding of NFTs" is gauged at 4, bearing an average of 3.45. This metric, if measured against fundamental themes like "Understanding of blockchain" (median: 4, average: 3.69) and "Understanding of bitcoin" (median: 4, average: 3.82), suggests that NFT knowledge trails only marginally. Given the visibility of NFTs within the crypto sector, this is an important observation. The predominant response for "Understanding of NFTs" is also 5, underscoring a robust knowledge level amidst a significant tranche of respondents.

Furthermore, tacit knowledge garnered by crypto-shoppers through their engagement in blockchain-enabled platforms and utilities is evident. Metrics such as "Participation in DAOs" (median: 3, average: 2.73) and "Playing blockchain-based games" (median: 3, average: 2.83) are noteworthy. Specifically, "Knowledge on using DeFi services" reveals a median of 3 with an average of 3.09, signifying a predominant familiarity with decentralized finance (DeFi) modalities in contrast to other blockchain applications.

\subsection{Findings on RQ2: Correlation of Knowledge and Expertise with Purchase Frequencies}
While the Cryptorefills Labs Reports \cite{Cabuk2021,Cabuk2022} abstain from analyzing the interplay between levels of knowledge, expertise, and their implications for crypto-shopping, this study harnesses both correlation analysis and regression modeling to furnish quantitative revelations of the dynamics of this relationship.

Our analysis reveals the presence of slight negative correlations spanning all knowledge facets and purchase frequency, most pronounced in the "Knowledge on Obtaining Bitcoin" (-0.11) metric. Such a trend suggests that amplification in domain expertise, especially in the realm of Bitcoin procurement, correlates with a marginal decrement in purchase frequencies. It is important to underscore, however, that notwithstanding their statistical validity, the strength of these correlations remains tenuous. The regression model suggests that knowledge dimensions explain a mere 11.6\% of the variability inherent in purchasing frequency, pointing to other factors not covered in this research.

Within the spectrum of the 17 examined variables, the act of obtaining Bitcoin emerges as the sole predictor exerting statistical significance upon purchase frequency. Interestingly, the model delineates that increments in knowledge concerning Bitcoin procurement are met with a decrement in purchase frequency by an approximate magnitude of 0.2981 units. This observation has implications, particularly in the context of Satoshi Nakamoto's foundational vision of Bitcoin as a digital currency paradigm. It potentially suggests that crypto-shoppers with a deeper experience in Bitcoin acquisition might align with the increasingly common perception of Bitcoin as a store of value or long-term investment rather than a means of exchange, thereby showing stronger reluctance in its expenditure.

While other knowledge areas did not strongly relate to purchasing habits, it's clear that while there's some link between specific knowledge and buying decisions, it's limited in how much it predicts behavior. 

Despite the appreciable knowledge levels of crypto-shoppers, the nexus between such knowledge and purchasing frequency remains nebulous. The analysis shows that domain expertise cannot be defined as the dominant precursor of crypto-shopping frequency, thereby implying the potential for a diverse cohort, spanning seasoned experts to new entrants or simply less knowledgeable cryptocurrency adopters, to engage in frequent crypto transactions irrespective of their knowledge depth.

\begin{figure}[!tb]
 \centering
 \includegraphics[scale=.15]{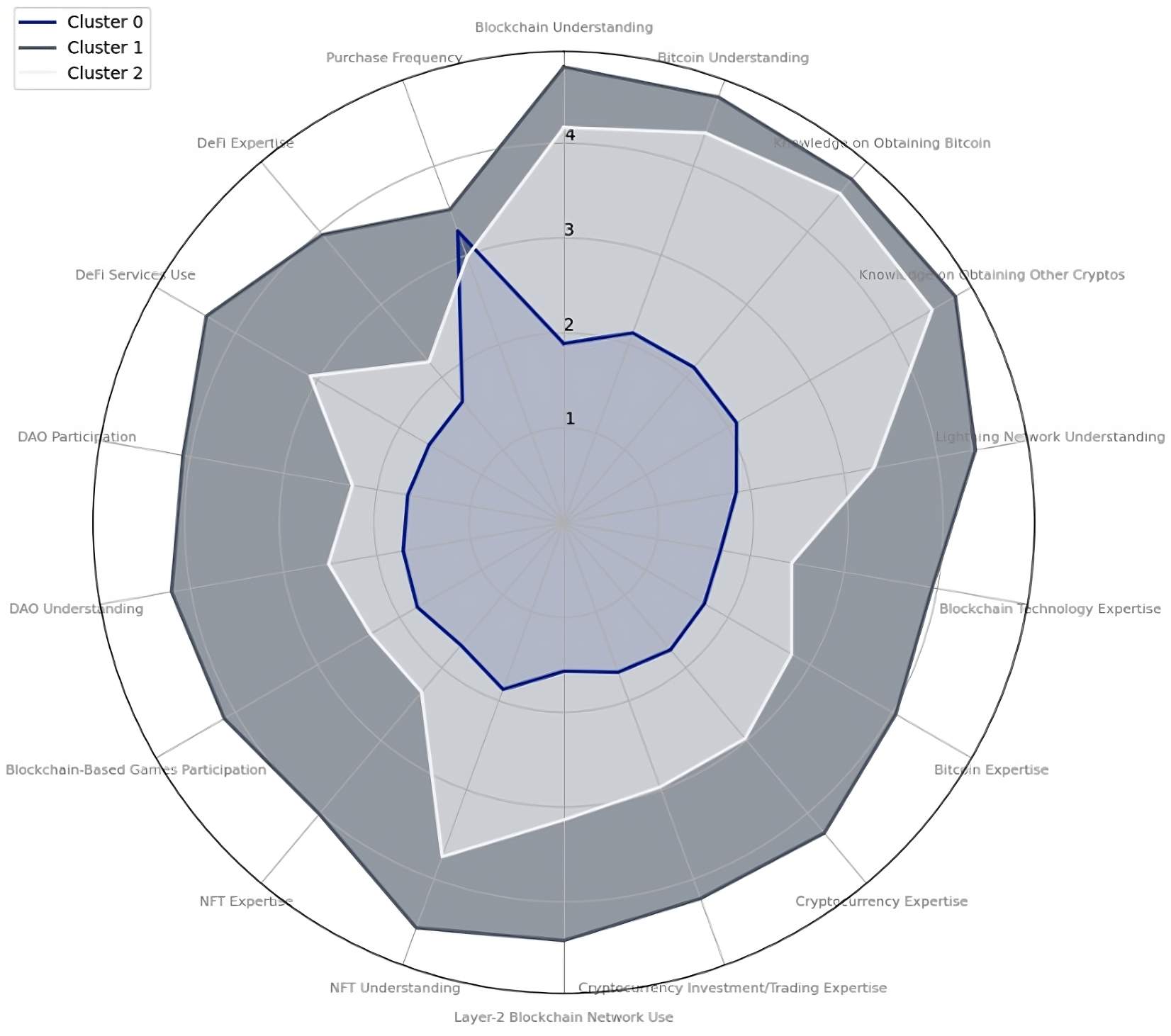}
 \caption{Radar map describing levels of knowledge, expertise, and purchase frequency by crypto-shopper segments.}
 \label{fig:fig6}
\end{figure}

\subsection{Findings on RQ3: Knowledge and Expertise-based Consumer Segmentation}
Throughout the years, the cryptocurrency sector has drawn many users, each with varying degrees of knowledge and expertise, especially among those who utilize cryptocurrencies for shopping. The present analytical research sought to meticulously dissect this demographic, emphasizing their knowledge, expertise, and purchase frequency vectors primarily. The main goal was to define and describe distinct clusters of crypto-shoppers predicated on their knowledge, expertise, and purchasing behaviors.

The application of the K-Means clustering model unveiled three crypto-shopper clusters. Cluster 0, labeled "The Low Knowledge Group", constituting 28.09\% of the populace, predominantly reflected low knowledge attributes. Conversely, Cluster 1, christened "The Experts" and accounting for 35.05\%, showed profound knowledge and expertise, encompassing both the more general crypto paradigms and domain-specific niches such as DAOs, NFTs, and DeFi. Cluster 2, the "Moderate Knowledge Group" at 36.85\%, delineated an intermediate-knowledge segment with pronounced foundational knowledge but relatively attenuated expertise in sectors like DeFi and DAOs. ANOVA outcomes validate the cluster distinctions, with a preponderance of the variables registering significant F-values and p-values < 0.001, accentuating the rigorousness of cluster demarcation. Characteristic features and differences of the groups are visualized in Figure \ref{fig:fig6}.

A deep dive into the Low Knowledge Group shows interesting dynamics. Despite their low or superficial blockchain and Bitcoin awareness, their crypto market participation is astoundingly vibrant, mirroring the Expert Group's purchase frequency. The observations may imply several interpretations. Firstly, the threshold for engaging in the crypto-shopping sphere may be diminishing, enabling individuals with minimal knowledge to actively partake. Secondly, the incentives for this demographic to utilize crypto for shopping could be predominantly influenced by external determinants—ranging from financial advantages and peer influence to the novelty of contemporary technologies—rather than a comprehensive grasp of the underlying tech. This behavioral trend aligns with Alqaryouti et al. \cite{Alqaryouti2020} that many users value cryptocurrency for its transactional merits over its technical foundations. The frequent transactions, despite constrained expertise, indicate that the appeal of features like decentralization and cost-effectiveness in crypto might outweigh the necessity for in-depth comprehension. The Expert Group's pronounced self-assessed knowledge and robust purchase frequency juxtapose interestingly against the Low Knowledge Group, highlighting that knowledge alone doesn't steer purchase decisions. The Moderate Knowledge Group, meanwhile, registers a subdued purchase frequency, perhaps denoting enthusiasm tempered with caution. 

Despite the statistically small variances in crypto-shopping frequency across clusters, their practical ramifications are palpable. For instance, the Moderate Knowledge Group's score of 2.98 translates into monthly shopping, while the Expert Group, with a score of 3.51, ventures into purchases multiple times monthly, sporadically even weekly. This seemingly subtle statistical distinction translates into a substantial divergence in real-world shopping habits, bridging the continuum from monthly to near-weekly crypto shopping.

\subsection{Business \& Managerial Implications}
This research delineates a sophisticated characterization of crypto shoppers, highlighting distinct patterns in terms of knowledge, expertise, and shopping behaviors, which hold implications for businesses contemplating or already offering cryptocurrency payments to their customers. An exhaustive examination of the findings lays the groundwork for potential strategies and spheres of interest that have the potential to increase growth, foster deeper customer engagement, and drive innovation in the cryptocurrency payments sector.

\begin{itemize}[itemsep=0.3em, parsep=0pt]
    \item \textbf{Expanding Acceptance:} The data reveals that the “Low Knowledge Users” demonstrate a noteworthy purchasing frequency, averaging once a month or more, suggesting a substantial market avenue for businesses willing to accept cryptocurrency payments. Such findings are congruent with the Cryptorefills Labs reports \cite{Cabuk2021,Cabuk2022}, which underscore specific demographics in fragile economic territories grappling with issues such as high inflation or marked financial instability. Despite their ostensibly limited expertise, the propensity for these “Low Knowledge Users” to engage in frequent transactions suggests that the barriers to entry in the crypto-commerce realm are potentially diminishing. This is further supported by the emergence of user-friendly technologies, like next-generation wallets, and a surging emphasis on superior user experiences and dedicated educational initiatives by leading industry players.
    \item \textbf{Segmented Marketing and Education:} The clear segmentation of crypto-consumers, founded on data-driven knowledge tiers, indicates an interesting opportunity for businesses offering crypto as a payment method to craft tailor-made marketing and educational campaigns. As an example, the segment termed “The Experts”, which presumably would encompass individuals boasting expertise in areas like blockchain gaming (35.27\%) and DAO participation (31.78\%), might be inclined to seek out advanced tools and services. In contrast, the "Low Knowledge Users" segment could find immense value in more streamlined interfaces coupled with content that accentuates the tangible, practical benefits inherent to crypto transactions.
    \item \textbf{Marketing Acquisition Channels:} From the data, it's evident that a good number of respondents have engaged in specific crypto-associated activities, with 28.29\% indicating expertise in DeFi, 29.84\% in NFTs, and a prominent cohort participating in blockchain gaming and DAOs. Businesses accepting cryptocurrency payments can, therefore, craft targeted partnerships, promotions, and advertising campaigns targeting the above channels to entice and acquire customers more inclined to transact with a cryptocurrency.
    \item \textbf{Pricing \& Payment Customization:} The data suggests that businesses accepting cryptocurrency can enhance user experience by customizing pricing and payment in relation to the demonstrated expertise levels of users. This can be implemented through innovative layer-2 protocol \cite{Zhou2020} offerings for the more seasoned "Connoisseurs", while for the "Practical Users", businesses can provide more direct, user-friendly options like simplified or even off-chain payments from accounts held at major crypto exchanges or on popular wallets.
    \item \textbf{Re-assessing Knowledge Assumptions:} A key takeaway from the data is the active transactional behavior of the “Low Knowledge Users”, challenging prevailing paradigms that link intricate domain knowledge directly with increased adoption. This indicates that businesses ought to be cautious not to deter a portion of their clientele by setting overly high knowledge expectations for crypto transactions.
    \item \textbf{Enhancing Trust:} The varying expertise observed in the data, particularly in segments such as "The Moderate Knowledge Group" and "The Practical Users", underlines the importance for businesses to emphasize security and build trust. This may involve sharing clear security guidelines, creating easy-to-follow guides, and promoting a transparent approach to operations.
\end{itemize}

\subsection{Limitations of Study}
As with all scholarly research, this study is not without its constraints. While our sample boasts a commendable margin of error under 5\% (at a 95\% CI), the elicited responses might not wholly reflect the knowledge strata and actual purchasing patterns amongst crypto-shoppers. Other limitations include:

\begin{itemize}[itemsep=0.3em, parsep=0pt]
    \item \textbf{Self-reported Data:} The research leans heavily on self-reported datasets, which can sometimes be influenced by recall bias, social desirability bias, or misinterpretations of questions. Yet, the self-perceived expertise may not align entirely with objective measures of knowledge.
    \item \textbf{Incentive Bias:}  Although the incentive was carefully designed to be appealing without exerting undue influence, there's always a possibility that it might have subtly influenced participation rates and the integrity of provided responses.
    \item \textbf{Geographical and Demographic Bias:} Our dataset is primarily derived from Cryptorefills clientele, which may not paint a holistic portrait of the global crypto-shopping population. The study might have inherent biases based on the platform's user demographics and geographical reach.
    \item \textbf{Cross-sectional Data:} The data is cross-sectional, taken at a specific point in time. Cryptocurrency is a rapidly evolving domain, and behaviors and knowledge levels might change over time. Longitudinal studies could offer deeper insights into how these aspects evolve.
    \item \textbf{CAWI Limitations:} Employing the computer-assisted web interviewing strategy, while operationally efficient, might inadvertently sideline potential respondents. Those who prefer alternative survey modalities or those besieged with technical impediments could be left out.
    \item \textbf{Potential Non-response Bias:} Not all solicited participants might have engaged with our survey. This could lead to inherent differences between respondents and non-respondents that might influence the results.
    \item \textbf{Single Platform Data Source:} Our database stems exclusively from Cryptorefills. While this imparts a degree of data consistency, it also introduces the caveat that behaviors and knowledge gradients on this platform might deviate from patterns observable on alternative platforms.
\end{itemize}

\subsection{Future Research}
The results of this study have shed light on several aspects of crypto-shoppers' knowledge and behavior. However, as with all research endeavors, new questions arise even as old ones find their answers. In this vein, we identify several promising avenues for future research. 

Firstly, this study primarily focuses on individuals who already shop with crypto. Yet, it's essential to recognize that these crypto shoppers represent only a fraction of the estimated 420 million crypto users globally. A logical progression would be to investigate the knowledge and expertise levels of a broader spectrum of crypto users. This could encompass those who hold cryptocurrencies but refrain from shopping with them, those familiar with the concept but have not ventured into its acquisition and even the general populace. Such a survey may offer richer insights into the role of knowledge and expertise in crypto adoption.

Another potential area of investigation lies in examining how knowledge interfaces with other determinants of behavior. While this study dives deep into the realm of knowledge, understanding its interplay with factors such as perceived ease of use, usefulness, trust, and perceived risks could paint a more holistic picture. This would provide a comprehensive view of the myriad factors influencing crypto shopping dynamics. The study expanded the knowledge and expertise variables to 17, providing a granular understanding. Yet, employing techniques like factor analysis could help consolidate these into fewer, more overarching constructs. 

From a theoretical standpoint, our empirical findings could be analyzed under established frameworks in the realm of technology adoption. Models like the Technology Acceptance Model and alike might find both resonance and divergence with the results in the specific context of cryptocurrencies. Such a theoretical exploration could further academic discourse in the domain.

Considering the rapidly evolving nature of the cryptocurrency space, a longitudinal approach could be particularly enlightening. Tracking knowledge, expertise, and behaviors over time would capture the dynamism inherent in this digital landscape, offering a temporal perspective on its evolution. Beyond quantitative surveys, qualitative endeavors such as in-depth interviews could offer richer insights. Personal narratives, motivations, hesitations, and experiences could provide the depth and context often elusive in numeric data. Furthermore, the principles of behavioral economics, with its focus on cognitive biases and heuristics, could be a lens to apply to crypto behaviors. By probing into the rational and often irrational patterns of crypto adoption and usage, we may uncover layers of psychological complexities at play.

Lastly, in a sub-domain of digital finances, trust is paramount. Future research can investigate how trust is established, maintained, or broken in the crypto world. Factors like security perceptions, past experiences with scams or hacks, and the role of crypto platforms could be pivotal in this exploration.

\section{Conclusions} \label{sec:conclusion}
Within the realm of cryptocurrency, characterized by its complex economic, societal and political aspects as well as its relentless adoption and technological progression, we provide a comprehensive delineation of blockchain domain knowledge and expertise among crypto-shoppers. These consumers, who actively engage in commercial transactions utilizing cryptocurrency, also display distinctive purchasing patterns, particularly concerning transaction frequency. Building upon prior foundational research, this study endeavors to define and describe these attributes.

Our research outcomes provide an increased comprehension of the crypto-shopper demographic, focusing on their expertise and understanding levels. A substantial majority of crypto-shoppers, approximately 70\%, exhibit a foundational understanding of the blockchain and cryptocurrency sectors. Yet, only a third of this populace confidently considers themselves experts. Numerous crypto-shoppers are deeply involved within the cryptocurrency ecosystem, leveraging DeFi platforms, engaging in blockchain-based games, or actively participating in DAOs. Such engagements could potentially unveil new marketing approaches for enterprises integrating cryptocurrency payment methodologies. Equally compelling is the revelation that a third of crypto-shoppers do not acknowledge possessing even a rudimentary understanding of blockchain, bitcoin, or other cryptocurrencies. This might suggest that the cryptocurrency adoption curve is expanding to encompass broader demographic cohorts, potentially propelled by practical needs superseding mere technological enthusiasm, including aspirations for financial inclusivity.

Furthermore, our study investigates the relationship between domain expertise and transactional frequency. Data derived from correlation and regression models demonstrate that profound expertise in the cryptocurrency arena is not a prerequisite for active participation. Interestingly, heightened expertise does not correlate with an uptick in transactional frequency. Contrarily, a slight negative correlation emerges, particularly when pertaining to knowledge of Bitcoin acquisition. This could allude to Bitcoin's utilization as a savings instrument or a store of value, given its stature as the premier cryptocurrency.

Segmenting our respondents into discrete clusters, spanning those with negligible knowledge to proclaimed experts, accentuates the diverse expertise spectra and behaviors intrinsic to crypto-shoppers. Every segment, with its characteristic traits, accentuates the heterogeneity of crypto-shopper personas. Such segmentation could proffer invaluable insights for businesses integrating cryptocurrency payment gateways, facilitating tailored marketing strategies, and optimizing user experience predicated on consumer expertise levels.

Nevertheless, while our discoveries significantly enhance the understanding of the crypto-shopper persona, it remains imperative to acknowledge inherent research limitations. The snapshot provided by this study captures a glimpse of the current situation in the ever-changing cryptocurrency environment, highlighting the need for ongoing observation and regular review.

\section{Acknowledgments}
This research is funded by Cryptorefills Labs and makes use of data collected from surveys offered to and responded to by Cryptorefills customers. The data used in this research is considered commercial and is not made available to the public. The authors acknowledge the use of AI-based text generator tools in this paper for text editing and grammar-checking purposes. 

\section{Glossary}
\begin{itemize}[label={}, itemsep=0.3em, parsep=0pt,wide=-0.5em]
   \item \textbf{Computer-Assisted Web Interviewing (CAWI)} is a surveying technique where participants complete questionnaires online, often sent via email or accessed through a website, without needing an interviewer's presence.
   \item \textbf{Cryptocurrency}, or informally crypto, is a digital form of currency that uses cryptography for security and operates independently of a central authority or traditional banking system. It serves as a decentralized medium of exchange, enabling direct peer-to-peer transactions over the Internet.
   \item \textbf{Crypto-shopper}, or crypto-consumer, refers to an individual who regularly purchases goods or services using cryptocurrencies as a payment method.
   \item \textbf{Decentralized Autonomous Organization (DAO)} is an organization represented by rules encoded as a transparent computer program, administered by the organization members, and not influenced by a central government. A DAO's financial transaction record and program rules are maintained on a blockchain.
   \item \textbf{Decentralized Finance (DeFi)} refers to a set of blockchain-based financial applications and protocols that aim to recreate traditional financial systems (e.g., lending, borrowing, and trading) without intermediaries, using smart contracts.
   \item \textbf{Layer-2 Blockchain Network} is a secondary protocol built atop an existing blockchain to enhance its scalability, efficiency, and speed. It processes transactions off-chain before settling them on the primary blockchain, reducing network traffic.
   \item \textbf{Non-fungible Token (NFT)} is a non-interchangeable data unit stored on a blockchain. Types of NFT data units may include digital files such as photos, videos, and audio. Since each token is uniquely identifiable, NFTs differ from cryptocurrencies.
    \item \textbf{Technology Acceptance Model 2 (TAM2)} is an extension of the Technology Acceptance Model (TAM) that incorporates perceived usefulness, perceived ease of use, and additional factors to predict information technology acceptance and usage.
\end{itemize}

\balance

\section{References} \vspace{-1.5em}
\bibliography{rsc} 
\bibliographystyle{rsc} 
\end{document}